\documentclass[aip,jmp,amsmath,amssymb,reprint]{revtex4-1}

\usepackage{empheq}
\usepackage{graphicx}
\usepackage{dcolumn}
\usepackage{bm}

\usepackage{cprotect}
\newcommand*{\citen}[1]{%
  \begingroup
    \romannumeral-`\x 
    \setcitestyle{numbers}%
    \cite{#1}%
  \endgroup   
}
\usepackage{soul,color}
\usepackage{etoolbox}
\usepackage{floatrow}

\begin{document}

\title{Apparent remote synchronization of amplitudes: a demodulation and interference effect}

\author{Ludovico Minati}
\altaffiliation{Author to whom correspondence should be addressed. Electronic addresses: ludovico.minati@ifj.edu.pl and lminati@ieee.org. Tel.: +39 335 486 670. URL: http://www.lminati.it.}
\affiliation{Complex Systems Theory Department, Institute of Nuclear Physics - Polish Academy of Sciences (IFJ-PAN), 31-342 Krak\'ow, Poland}
\affiliation{Center for Mind/Brain Sciences (CIMeC), University of Trento, 38123 Trento, Italy}
\author{Luca Faes}
\affiliation{Department of Energy, Information Engineering and Mathematical Models (DEIM), University of Palermo, 90128 Palermo, Italy}
\affiliation{BIOTech - Department of Industrial Engineering, University of Trento, 38123 Trento, Italy}
\author{Mattia Frasca}
\affiliation{Department of Electrical Electronic and Computer Engineering (DIEEI), University of Catania, 95131 Catania, Italy}
\author{Pawe\l\ O\'{s}wi\c{e}cimka}
\affiliation{Complex Systems Theory Department, Institute of Nuclear Physics - Polish Academy of Sciences (IFJ-PAN), 31-342 Krak\'ow, Poland}
\author{Stanis\l aw Dro\.{z}d\.{z}}
\affiliation{Complex Systems Theory Department, Institute of Nuclear Physics - Polish Academy of Sciences (IFJ-PAN), 31-342 Krak\'ow, Poland}
\affiliation{Faculty of Physics, Mathematics and Computer Science, Cracow University of Technology, 31-155 Krak\'ow, Poland}
\date{\today}
\begin{abstract}
A form of ``remote synchronization'' was recently described wherein amplitude fluctuations across a ring of non-identical, non-linear electronic oscillators become entrained into spatially-structured patterns. According to linear models and mutual information, synchronization and causality dip at a certain distance, then recover before eventually fading. Here, the underlying mechanism is finally elucidated through novel experiments and simulations. The system non-linearity is found to have a dual role: it supports chaotic dynamics, and it enables energy exchange between the lower and higher sidebands of a predominant frequency. This frequency acts as carrier signal in an arrangement resembling standard amplitude modulation, wherein the lower sideband and the demodulated baseband signals spectrally overlap. Due to a spatially-dependent phase relationship, at a certain distance near-complete destructive interference occurs between them, causing the observed dip. Methods suitable for detecting non-trivial entrainment, such as transfer entropy and the auxiliary system approach, nevertheless reveal that synchronization and causality actually decrease with distance monotonically. Remoteness is, therefore, arguably only apparent, as also reflected in the propagation of external perturbations. These results demonstrate a complex mechanism of dynamical interdependence, and exemplify how it can lead to incorrectly inferring synchronization and causality.
\end{abstract}
\maketitle

\begin{quotation}
There are diverse mechanisms through which dynamics can give rise to complex synchronization patterns between non-linear units, having features potentially absent in the underlying structural connections. One such mechanism is ``remote synchronization'', namely the selective or preferential entrainment between groups of units (e.g., oscillators) that are not structurally connected. This may be observed, for example, between neurons in distant brain regions or meteorological processes over different continents. Here, we present an explanation for a peculiar form of remote synchronization, previously discovered in a ring of electronic non-linear oscillators. As the units are parametrically heterogeneous, identical synchronization is unlikely to occur. Yet, in this system the intensity of synchronization of amplitude fluctuations appears to dip at a certain distance along the ring, and then recover before eventually fading. By means of new experiments and simulations, it is found that this dip reflects an interference phenomenon, which is more closely related to ``local'' interactions than to the collective behavior of the network as a whole. The effect arises in a scenario reminiscent of amplitude modulation, wherein the system non-linearity demodulates the envelope signal. Importantly, it is demonstrated that the dip is ultimately due to choosing unsuitable synchronization and causality measures. Techniques capable of detecting sufficiently generalized interdependencies show that actually the intensity of both decreases monotonically with distance. These results prompt further reflection on the definition and measurement of remoteness. 
\end{quotation}

\section{INTRODUCTION}\label{intro}
Coupled non-linear oscillators have an extraordinary generative potential with respect to forming synchronization patterns not straightforwardly related to structural connections\cite{arenas2008synchronization,pecora2014cluster,fischer2006zero,gambuzza2016inhomogeneity,abrams2004chimera,panaggio2015chimera,bergner2012remote,gambuzza2013analysis,nicosia2013remote}. While the means through which such patterns can be formed are diverse and difficult to classify univocally, well-known phenomena include cluster synchronization\cite{pecora2014cluster}, relay synchronization\cite{fischer2006zero,gambuzza2016inhomogeneity}, formation of chimera states\cite{abrams2004chimera,panaggio2015chimera} and remote synchronization \cite{bergner2012remote,gambuzza2013analysis,nicosia2013remote}. The latter encompasses heterogeneous mechanisms, which lead to subsets of non structurally-connected nodes becoming more strongly synchronized than the intermediate relay nodes, or even becoming selectively synchronized. One striking example is the identical (complete) synchronization observed in laser systems with delayed coupling\cite{fischer2006zero}.\\
Known instances of remote synchronization may be broadly grouped according to whether they involve parametric mismatches\cite{bergner2012remote,gambuzza2013analysis} or identical nodes\cite{nicosia2013remote,zhang2017incoherence}. As regards the first group, the term remote synchronization has been used to denote the onset of phase synchronization among the outer nodes of a star network of periodic oscillators, wherein the central hub has mismatched natural oscillation frequency and is not synchronized with them\cite{bergner2012remote}. This phenomenon partially overlaps that of relay synchronization, which is observed for chaotic oscillators and manifests as complete synchronization of the outer nodes mediated by the central one, which is synchronized with them in a weaker sense, such as with a lag\cite{fischer2006zero} or a generalized relationship\cite{gutierrez2013generalized}. Notably, both phenomena are also found in complex random topologies beyond the simple motifs where they have first been observed \cite{gambuzza2013analysis,gambuzza2016inhomogeneity}. As regards the second group, symmetries in the structural connections engender the emergence of remotely-synchronized node subsets. In this context, the term remote synchronization has been used to refer to the formation of node subsets having coherent phase or common trajectory, and interconnected via other nodes involving either a lag\cite{nicosia2013remote} or completely incoherent\cite{zhang2017incoherence} dynamics.\\
Given heterogeneous elements, identical synchronization is largely precluded in ring networks. Nevertheless, one may observe rich repertoires of dynamical behaviors, transitions among them and variability of spatio-temporal patterns, which are of fundamental importance also in theoretical neuroscience \cite{perlikowski2010routes,yanchuk2011variability,d2011role}. In Ref. \citen{minati2015}, a peculiar form of remote synchronization was elicited in a ring of electronic non-linear oscillators involving small parametric mismatches. It manifested as a non-monotonic decay of the synchronization of amplitude fluctuations with distance, and lead to the appearance of patterns having complex features such as small-world topology. The effect was extensively demonstrated through simulations and an experimental implementation based on field programmable analog arrays (FPAAs). It was suggested that it involves the formation of interdependencies between coupled units, which are visible only to suitably generalized synchronization measures. However, a true understanding of the underlying mechanism was not attained.\\
Here, a detailed account is finally given. After a description of the system structure and observed synchronization effect, the interactions between oscillators are revisited from an information-theoretical perspective. This highlights that remote synchronization is only apparent insofar as inadequate synchronization and causality measures are considered. Further, new experiments involving external perturbation with white noise and sinc pulses are performed. Alongside simulations of different-size rings, an auxiliary system and a simplified chain model, these experiments unequivocally demonstrate that the effect does not represent a global, collective behavior of the network. Rather, it reflects complicated local interactions between each node and its surroundings.

\begin{figure*}
\floatbox[{\capbeside\thisfloatsetup{capbesideposition={right,center},capbesidewidth=5cm}}]{figure}[\FBwidth]
{\caption{Chaotic oscillator network. a) Circuit instanced at each node, consisting of a ring oscillator to which two integrators are superimposed. b) Network comprising 32 such circuits forming a ring via master-slave coupling. See Subsection \ref{basics1} for detailed description.\label{fig:fig1}}}
{\includegraphics[width=0.65\textwidth]{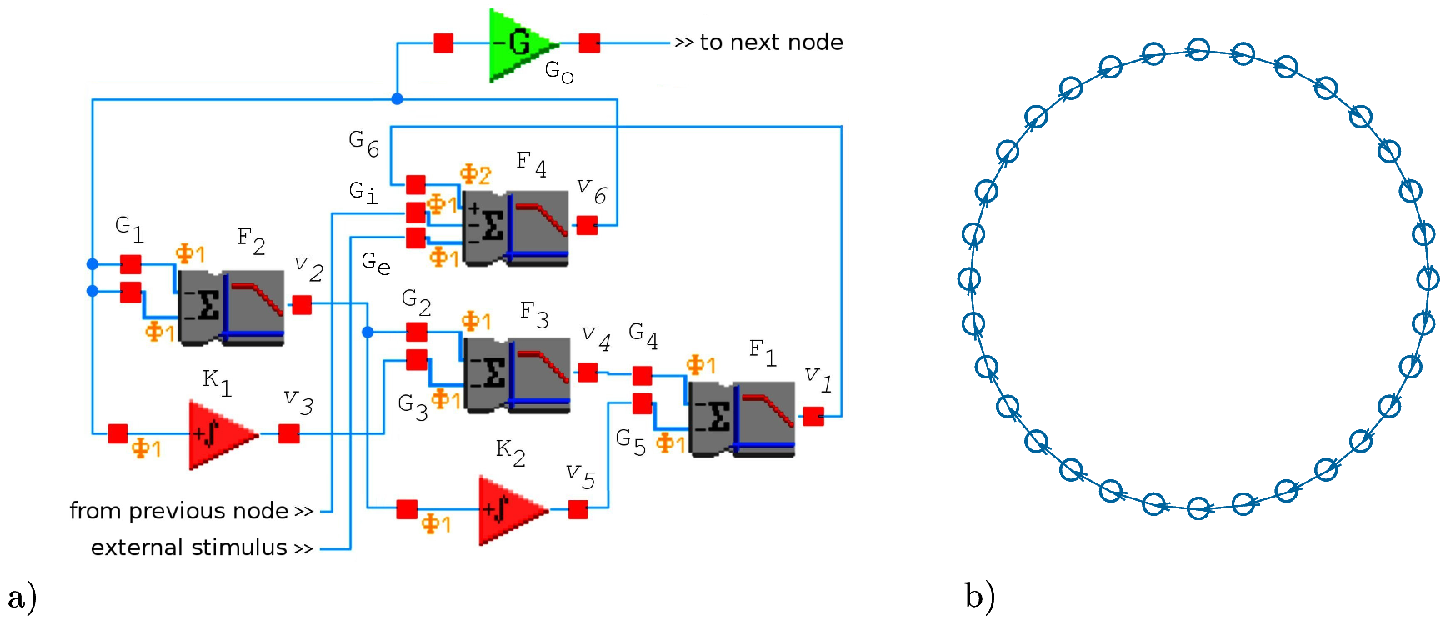}}
\end{figure*}

\begin{figure*}
\centering
\includegraphics[width=0.9\textwidth]{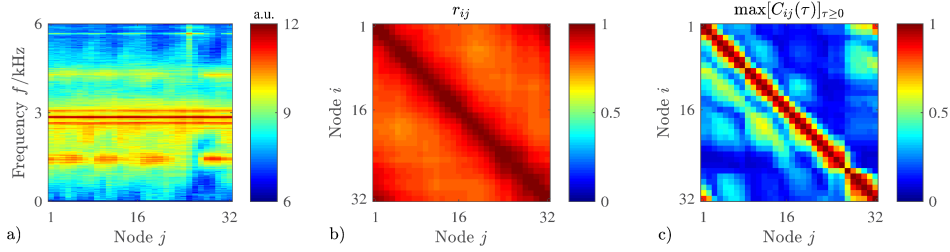}
\cprotect\caption{Basic features of spontaneous activity for a representative, experimentally-measured signal in the remote synchronization region. a) Spectrogram depicting the Fourier log-amplitudes across the nodes. b) Phase coherence. c) Maximum positive-lag cross-correlation coefficient for the signal envelope (amplitude fluctuations, thereafter referred to simply as amplitude synchronization). Raw signals from Ref. \citen{minati2015}. See Subsection \ref{basics2} for detailed description.\label{fig:fig2}}
\end{figure*}

\begin{figure*}
\floatbox[{\capbeside\thisfloatsetup{capbesideposition={right,center},capbesidewidth=5cm}}]{figure}[\FBwidth]
{\caption{Mutual information, causality and transfer entropy. a) Normalized mutual information, calculated based on the scalar time-series. b) Granger causality according to the linear model. c) Granger causality according to a model including quadratic terms and cross-terms. d) Transfer entropy, calculated with a model-free approach. See Section \ref{causality} for detailed description.\label{fig:fig3}}}
{\includegraphics[width=0.6\textwidth]{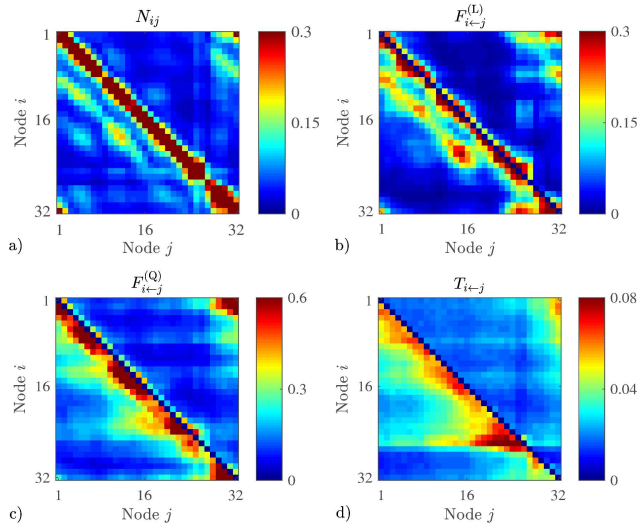}}
\end{figure*}
\section{SYSTEM STRUCTURE AND DYNAMICS}\label{basics}
\subsection{System specification}\label{basics1}
Each oscillator consists of a ring of three composite stages which implement gain, summation and low-pass filtering, and of two integrators whose outputs enter different points of the ring (Fig. \ref{fig:fig1}a). This structure effectively forms a sort of ring oscillator, and was developed specifically since it supports rich dynamics, and at the same time is suitable for physical deployment using a single FPAA device\cite{minati2015}.\\
The oscillator dynamics are governed by the following system
\begin{empheq}[left=\empheqlbrace]{align}
\frac{\textrm{d}v_{1}}{\textrm{d}t}&=\Gamma\Big(2\pi F_1(G_4v_4+G_5v_5-v_1),v_1\Big)\nonumber\\
\frac{\textrm{d}v_{2}}{\textrm{d}t}&=\Gamma\Big(2\pi F_2(G_1v_6-v_2),v_2\Big)\nonumber\\
\frac{\textrm{d}v_{3}}{\textrm{d}t}&=\Gamma\Big(K_1v_6,v_3\Big)\nonumber\\
\frac{\textrm{d}v_{4}}{\textrm{d}t}&=\Gamma\Big(2\pi F_3(G_2v_2+G_3v_3-v_4),v_4\Big)\\
\frac{\textrm{d}v_{5}}{\textrm{d}t}&=\Gamma\Big(K_2v_2,v_5\Big)\nonumber\\
\frac{\textrm{d}v_{6}}{\textrm{d}t}&=\Gamma\Big(2\pi F_4(G_6v_1+G_{\textrm{i}}v_{\textrm{i}}+G_{\textrm{e}}v_{\textrm{e}}-v_6),v_6\Big)\nonumber
\end{empheq}
wherein the only non-linearity is the function $\Gamma(x,y)$, which represents an approximation of saturation effects due to finite voltage swing $V_{\textrm{s}}$ according to
\begin{eqnarray}
\Gamma\left(x,y\right)=\textrm{R}\left(x\right)\textrm{H}(V_{\textrm{s}}-y)-\textrm{R}\left(-x\right)\textrm{H}(V_{\textrm{s}}+y)
\end{eqnarray}
where the Heaviside step function $\textrm{H}(x)=1\textrm{ for }x>0\textrm{, }0\textrm{ for }x\le0$ and the ramp function $\textrm{R}(x)=x\textrm{H}(x)$. The oscillator output voltage corresponds to $v_{\textrm{o}}=G_{\textrm{o}}v_6$, where $G_{\textrm{o}}=-0.4$.\\
Each oscillator receives as input the signal, denoted as  $v_{\textrm{i}}$, from the preceding node on a unidirectionally (e.g., master-slave) coupled ring of 32 oscillators (Fig. \ref{fig:fig1}b); in addition, $v_{\textrm{e}}$ denotes the external perturbation voltage which was applied during the experiments reported in Section \ref{perturbation}. As clarified below, a signal acting as a ``carrier'' is generated, and features a phase lead in the coupling direction engendered by the integrative dynamics. Concomitantly, large  amplitude fluctuations emerge, for the propagation of which a lag is observed in the direction of coupling.\\
Unless otherwise noted, we set $G_1=-3.6$, $G_2=-3.12$, $G_3=-0.5$, $G_4=-3.08$, $G_5=-0.71$, $G_6=0.132$, $F_1=F_2=F_3=2\textrm{ kHz}$, $F_4=100\textrm{ kHz}$, $K_1=3.67K_2=0.11\ \mu\textrm{s}^{-1}$, $G_{\textrm{i}}=-1.44$, $G_{\textrm{e}}=0$ and $V_{\textrm{s}}=4\textrm{ V}$, as for the case shown in Fig. 7a of Ref. \citen{minati2015}. We re-analyzed the corresponding experimental time-series, which are available in the Supplementary Materials of Ref. \citen{minati2015}, and also available for an extended set of parameter settings from Ref. \citen{minati2015web}. The recorded time-series have length $l=\textrm{65,536 points}$, are sampled at 8-bit and $f_\textrm{s}=100\textrm{ kSa/s}$, and no averaging over multiple runs is performed.\\
Due to reasons related to construction of the physical system, there are unavoidable mismatches in the realization of these parameters, which are on the order of 0.5\%. The values indicated above therefore prescribe the nominal values for the parameters. Given a parameter $P$ having average $\langle P\rangle$ (representing $G_1$, $G_2$ etc.), mismatches can be represented for example by extracting individual realizations from a uniform distribution of the form
\begin{equation}
f(x)=
\begin{cases}
\frac{1}{2\xi\langle P\rangle} & \text{for}\ \langle P\rangle(1-\xi)\le x\le\langle P\rangle(1+\xi) \\
0 & \text{otherwise}
\end{cases}
\end{equation}
where $\xi=0.005$. As detailed in Supplementary Section I, the observed form of remote synchronization has a large basin of attraction, emerging for $\xi\in[10^{-17},10^{-3}]$. Below this range, the oscillators are close to identical and it is replaced by global synchronization, whereas above it partial synchronization without remoteness ensues.
\subsection{Basic dynamical and synchronization features}\label{basics2}
The frequency spectrum features three broad but clearly-distinct peaks: a prominent central peak at $f_{\textrm{c}}\approx2.8\textrm{ kHz}$ (here appearing split, plausibly due to the parametric mismatches) and two weaker peaks at $f_{\textrm{l}}=f_{\textrm{c}}/2\approx1.4\textrm{ kHz}$ and $f_{\textrm{h}}=f_{\textrm{l}}+f_{\textrm{c}}\approx4.2\textrm{ kHz}$. These have the appearance of its lower and higher sidebands (Fig. \ref{fig:fig2}a), and faint resonances are also found at $f_{\textrm{l}}+f_{\textrm{h}}\approx5.6\textrm{ kHz}$ and $f_{\textrm{c}}+f_{\textrm{h}}\approx7.1\textrm{ kHz}$ (data not shown). The lower sideband reflects amplitude fluctuations, and specifically the relationship $f_{\textrm{l}}=f_{\textrm{c}}/2$ is verified because of strong anti-persistence, having Hurst exponents\cite{ihlen2012} $H_\textrm{max}=0.15\pm0.00$ and $H_\textrm{min}=0.26\pm0.00$ for the maxima and minima respectively. This stems from an underlying period-doubling bifurcation, which is demonstrated in Supplementary Section I, where it is also confirmed that $\lambda_\textrm{max}>0$. The higher sideband represents the mirror frequency of the lower one. As elucidated in Section \ref{interactionbands}, demodulation via envelope detection and subsequent interference occur. This leads to spatial fluctuations of the lower sideband amplitude, which resemble a diffraction pattern and are closely related to the remote synchronization effect (Fig. \ref{fig:fig2}a).\\
As in Ref. \citen{minati2015}, the instantaneous phase $\varphi_i(t)$ and amplitude $A_i(t)$ (i.e., envelope) of the output signal $v_i(t)$ of each oscillator $i$ were calculated via the analytic signal, according to
\begin{equation}
v_i(t)+\textrm{i}\hat{v}_i(t)=A_i(t)\textrm{e}^{\textrm{i}\varphi_i(t)}
\end{equation}
where $\hat{v}_i$ is the Hilbert transform of $v_i(t)$
\begin{equation}
\hat{v}_i(t)=\frac{1}{\pi}\textrm{p.v.}\left[\int_{-\infty}^\infty\frac{v_i(\tau)}{t-\tau}\textrm{d}\tau\right]
\end{equation}
and where $\textrm{p.v.}$ denotes the Cauchy principal value of the integral \cite{boashash1992}.\\
Phase coherence was measured with 
\begin{equation}
r_{ij}=|\langle\textrm{e}^{\textrm{i}[\varphi_i(t)-\varphi_j(t)]}\rangle| 
\end{equation}
and revealed global, albeit predictably imperfect, phase synchronization (Fig. \ref{fig:fig2}b).\\
Amplitude synchronization for $A_i(t)$ was considered in terms of the maximum normalized cross-correlation coefficient for non-negative lags, i.e.,  $\max[C_\textrm{XY}(\tau)]_{\tau\ge0}$, given
\begin{equation}
C_\textrm{XY}(\tau)=\frac{k_\textrm{XY}(\tau)}{\sqrt{\sigma_X^2\sigma_Y^2}}
\end{equation}
where $k_\textrm{XY}(\tau)=\mathbb{E}[(x_{n+\tau}-\mu_\textrm{X})(y_n-\mu_\textrm{Y})]$ is the time-lagged cross-covariance, and
$\sigma^2_\textrm{X}=\mathbb{E}[(x_n-\mu_\textrm{X})^2]$ and $\mu_\textrm{X}=\mathbb{E}[x_n]$ represent the variance and mean of time-series $\textrm{X}=x_n$.\\
It revealed a pattern wherein, besides incomplete regularity due to the parametric heterogeneity, the predominant feature was that, moving away from a given node, synchronization initially decayed, then increased again peaking at a distance $d=i-j\approx8$, and eventually vanished (Fig. \ref{fig:fig2}c). The corresponding ``synchronization dip'' could indicate a form of remote synchronization, because structural coupling on the ring is only between first neighbors, yet there appears to be stronger entrainment for some nodes at distance $d>5$ than at distance $d\approx5$.\\
This effect could be related to an underlying interdependence which is non-linear, such that $y(t)\not\propto x(t)$ and Eq. (7) loses meaning as a synchronization measure.\\
To test this hypothesis, amplitude synchronization was also assessed based on the maximum normalized time-lagged mutual information $N_\textrm{XY}(d)$ for non-negative lags $d\ge0$, defined as
\begin{equation}
N_\textrm{XY}(d)=\frac{I_\textrm{XY}(d)}{\sqrt{H_\textrm{X}H_\textrm{Y}}}
\end{equation}
where the Shannon entropies $H_\textrm{X}=-\mathbb{E}[\log_2 p(x_n)]$, $H_\textrm{Y}=-\mathbb{E}[\log_2 p(y_n)]$ and the time-lagged mutual information
$I_\textrm{XY}(d)=\mathbb{E}[\log_2 (p(x_{n+d},y_n) / (p(x_n)p(y_n)))]$
depend on the marginal probability distributions $p(x_n)$, $p(y_n)$ and on the time-lagged joint distribution $p(x_{n+d},y_n)$. The entropies and mutual information were estimated by histogram quantization using $Q=16$ bins\cite{paninski2003}. The maximal mutual information was searched for over a range of delays $d \in[0,2]$ ms, which covers the interval estimated for information propagation (larger delay ranges produced comparable results).\\
Notably, mutual information tracked cross-correlation very closely (Fig. \ref{fig:fig3}a), as confirmed by rank-order correlation between $\max[C_\textrm{XY}(\tau)]_{\tau\ge0}$ and $\max[N_\textrm{XY}(d)]_{d\in[0,2]}$, yielding $r=0.86$ and $p<0.001$. The failure of mutual information to reveal a monotonic synchronization decay recalls previous simulations and experiments on networks of Mackey-Glass oscillators. In those networks, a form of generalized synchronization ensued, which was visible to neither measure. However, when an auxiliary system was constructed, it revealed strong entrainment between the response and replica oscillators\cite{soriano2012}. As described in Sections \ref{causality} and \ref{ringsaux}, generalized synchronization also underlies the form of remote synchronization observed for the present system.
\section{LINEAR, QUADRATIC AND MODEL-FREE CAUSALITY ANALYSES}\label{causality}
\subsection{Rationale}\label{causality1}
Synchronization measures such as the maximum cross-correlation coefficient and mutual information are symmetric, i.e. given opposite lags, they are equal in the $i\rightarrow j$ and $j\rightarrow i$ directions. They are therefore inherently incapable of detecting causality between coupled units in an observational framework (referred to as \textit{information transfer}). This is exemplified by two identical oscillators having the same starting conditions, which remain perfectly synchronized in the absence of any energy exchange. Even observation of temporal precedence via time-lagged measures is inadequate for establishing causality, because it does not distinguish the information actually exchanged by the units from that, which is only shared in response to common inputs or past history\cite{schreiber2000,wibral2011}.\\
These issues are of particular concern since in the present system the oscillators are nearly identical, i.e. $\xi\approx0$, and the connectivity is closed-loop. For correctly inferring causality between two coupled units in an observational framework, meaning without probing the system response to an external perturbation, one needs to quantify the amount of information shared between the past states of the driver and the present state of the target, which is not accounted for the past states of the target. This is essential because deterministic systems such as chaotic oscillators can generate autocorrelated activity. As a consequence, they can store significant amounts of information, which confounds the relationship between information transfer and synchronization measures\cite{ottbook,lizier2012}.\\
We therefore investigated whether in the present system remote synchronization truly represents remote information transfer. To this end, we applied Granger causality computed in its parametric formulation based on linear and nonlinear regression models \cite{granger1969,geweke1982}. We also applied it in its nonparametric formulation directly expressed in information-theoretic terms and known as transfer entropy\cite{schreiber2000,wibral2011}. These measures of information transfer reflect causality according to a probabilistic notion, positing that activity in a node $j$ is Granger-causal to activity in another node $i$ if and only if knowledge of the past of node $j$ improves prediction of the present of node $i$ above and beyond the predictability yielded by knowledge of the past of node $i$ itself\cite{granger1969}. Since in the present system a form of remote synchronization is observed for signal envelopes, for these analyses we represented dynamics with the instantaneous amplitudes, i.e. $y_i(t)=A_i(t)$, as done in Subsection \ref{basics2} for cross-correlation and mutual information.
\subsection{Linear analysis}\label{causality2}
First, causality was investigated by performing a linear regression of the present state of each node $i$, $y_i(t)$, once given its past state $Y_i^-(t)$, yielding the prediction error $e_{i|i}(t)=y_{i}(t)-\mathbb{E}[y_{i}(t)|Y_i^-(t)]$, and once given the past states of both $i$ and $j$, $Y_{i,j}^-(t)=[Y_i^-(t),Y_j^-(t)]$, yielding prediction error $e_{i|i,j}(t)=y_{i}(t)-\mathbb{E}[y_{i}(t)|Y_{i,j}^-(t)]$.\\
Here, the past states of the target node were approximated as $Y_i^-(t)=[y_i(t-\delta), y_i(t-\tau-\delta), \ldots, y_i(t-p\tau-\delta)]$, i.e. using uniform embedding with lagged components equally spaced in time and introducing a term with lag $\delta=1/f_\textrm{s}=0.01 \textrm{ ms}$ to ensure full elimination of information storage\cite{wibral2013}. The past states of the drivers were similarly approximated as
$Y_j^-(t)=[y_j(t-\delta-d), y_j(t-\tau-\delta-d), \ldots, y_j(t-p\tau-\delta-d)]$, introducing however an additional lag $d$, to account for propagation delays. For each node pair $(i,j)$, the lag was chosen by searching for the minimal prediction error over the range $d\in[0,2]\textrm{ ms}$.\\
The prediction error variances given by these two regressions, $\lambda_{i|i}=\mathbb{E}[e_{i|i}^2]$ and $\lambda_{i|i,j}=\mathbb{E}[e_{i|i,j}^2]$, were combined to yield the Granger causality from node $j$ to node $i$\cite{geweke1982}
\begin{equation} 
F_{i\leftarrow j}^{\textrm{(L)}}=\log\frac{\lambda_{i|i}}{\lambda_{i|i,j}}.
\end{equation}
In this and the subsequent analyses, the lag $\tau$ was set to 0.18 ms, corresponding to the first local minimum of the time-delayed mutual information, taken as an empirical estimator of the time-scale of dynamics\cite{fraser1986}. Determination of $p$ was more problematic, given that the Akaike and Bayesian Information Criteria (AIC, BIC)\cite{ward2008} both decreased monotonically up to $p=20$, indicating that a linear model does not fully describe the dynamics. Ultimately, $p=16$ was chosen as a compromise sufficiently covering past history without excessively elevating model order, and $p=8\textrm{ or }32$ yielded comparable results. To reduce computational load, the time-series were decimated by a factor of 4.\\
As shown in Fig. \ref{fig:fig3}b, the resulting pattern closely tracked cross-correlation and mutual information (Fig. \ref{fig:fig2}c). This confirms that, insofar as a linear causality model is concerned, in this system the observed form of remote synchronization reflects remote information transfer, which was strongly asymmetric, as expected given the unidirectional coupling (Fig. \ref{fig:fig1}b).
\subsection{Non-linear and model-free analyses}\label{causality3}
However, it has previously been established that a linear model is insufficient to capture the dynamics of this system\cite{minati2015}. To overcome this limitation while retaining a parametric approach, the model was extended to account for non-linear Granger-causal influences by including quadratic terms (e.g., $y^2_i(t-\delta-m\tau)$) and cross-terms (e.g., $y_i(t-\delta-m\tau)y_i(t-\delta-n\tau)$ for $m\neq n$) into the separate regressors $Y_i^-(t)$ and $Y_j^-(t)$\cite{faes2008}. The resulting nonlinear measure is referred to as $F_{i\leftarrow j}^{\textrm{(Q)}}$. For consistency, the $\tau$, $p$ and $\delta$ settings were retained, alongside the range of $d$.\\
As shown in Fig. \ref{fig:fig3}c, we observed $F_{i\leftarrow j}^{\textrm{(Q)}}> F_{i\leftarrow j}^{\textrm{(L)}}$. Moreover, the causality pattern according to $F_{i\leftarrow j}^{\textrm{(Q)}}$ was substantially different from that of $F_{i\leftarrow j}^{\textrm{(L)}}$ in that it featured a gradual, essentially monotonic decay of information transfer with respect to distance. This result provides an initial explicit indication that causal interdependence in this system is actually not remote, simply it is ``hidden'' to the linear model for an intermediate range of distances, where a linear relationship fades but a more complex interdependence is maintained.\\
A more general measure of information transfer, namely transfer entropy, can be defined without recourse to any model, directly in probabilistic terms as a conditional mutual information\cite{schreiber2000}, according to
\begin{equation}
T_{i\leftarrow j}=I(y_{i}(t);Y_j^-(t)|Y_i^-(t))
\end{equation}
and practically computed by estimating its four entropy terms\cite{wibral2011}
\begin{equation}
\begin{split}
I(y_{i}(t);Y_j^-(t)|Y_i^-(t))=&H(y_{i}(t),Y_i^-(t))-H(Y_i^-(t))-\\
&H(y_{i}(t),Y_{i,j}^-(t))+H(Y_{i,j}^-(t))\textrm{ .}
\end{split}
\end{equation}
Here, $T_{i\leftarrow j}$ was estimated with the nearest-neighbor method, implemented with a distance projection which compensates for the different biases of the individual entropy estimates \cite{kraskov2004,faes2015b}. We set $k=10$ (this choice is not critical), and reduced the order to $p=8$ since for higher settings $T_{i\leftarrow j}$ dropped considerably due to the excessive estimation bias induced by the curse of dimensionality. All calculations were performed using the ITS toolbox available at Ref. \citen{faesits}, which implements the estimators described in Ref. \citen{faes2015b}.\\
As shown in Fig. \ref{fig:fig3}d, the pattern closely resembled that obtained from the non-linear Granger causality model, suggesting that the proposed non-linear model already captures the dynamics adequately\cite{barnett2009}, and confirming the results obtained with a rank-based measure of generalized synchronization in the initial study\cite{minati2015}. As will become clear in Supplementary Section II, this result can be recovered with an ad-hoc linear Granger model explicitly representing the overlapped low-frequency signal and demodulated baseband signal.

\begin{figure*}
\floatbox[{\capbeside\thisfloatsetup{capbesideposition={right,center},capbesidewidth=5cm}}]{figure}[\FBwidth]
{\caption{Perturbation by white noise. a) Representative signals measured before ($v_8$), at ($v_9$) and after ($v_{10}$, $v_{11}$) a site where high-intensity white noise was injected. b) Change in amplitude synchronization (maximum cross-correlation coefficient) due to noise injection, averaged across all nodes while rotating the ring to center the injection site $\Delta i=\Delta j=0$. c) and d) Corresponding examples for noise injection at nodes $j=9,17$. See Subsection \ref{perturbation1} for detailed description.\label{fig:fig4}}}
{\includegraphics[width=0.6\textwidth]{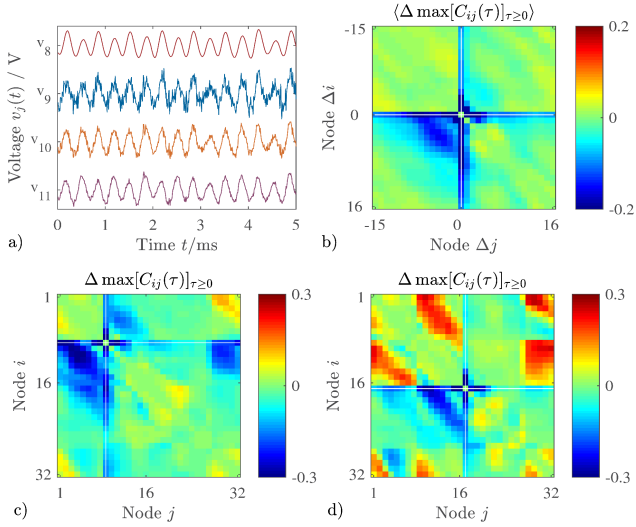}}
\end{figure*}

\begin{figure*}
\floatbox[{\capbeside\thisfloatsetup{capbesideposition={right,center},capbesidewidth=5cm}}]{figure}[\FBwidth]
{\caption{Propagation of sinc pulses. a) Amplitude synchronization averaged while injecting low-amplitude sinc pulses at node $j=4$, showing negligible difference compared to Fig. \ref{fig:fig2}c. b) Monotonic decay of the sinc pulse amplitude with distance along the ring (red, stars indicate statistically-significant signal detected), with concomitant remote synchronization of intrinsic activity at node $j=12$ (blue). c) and d) Time-locked signal average calculated for the injection site and the site corresponding to the remote synchronization peak, $j=4,12$. See Subsection \ref{perturbation2} for detailed description.\label{fig:fig5}}}
{\includegraphics[width=0.6\textwidth]{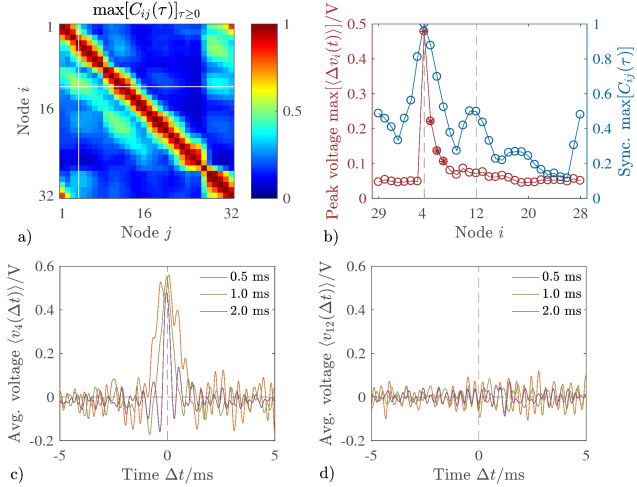}}
\end{figure*}

\section{PERTURBATION BY WHITE NOISE AND SINC PULSES}\label{perturbation}
\subsection{Rationale}
The observational framework considered thus far does not fully clarify whether the present form of remote synchronization is underpinned by local interaction between each node and its surroundings, a global phenomenon such as a collective oscillation mode, or perhaps a combination between the two. To address this issue conclusively, it is necessary to conduct new experiments involving external interventions, i.e. perturbing the intrinsic activity with unrelated stimuli\cite{pearl2003,lizier2010}.\\
The time-series from these additional experiments, which are not part of the initial study\cite{minati2015,minati2015web}, have also been made publicly available as a separate archive\cite{onlinemat}.
\subsection{Noise-induced desynchronization}\label{perturbation1}
It is well established that, by causing dispersion in phase space, additive noise perturbing individual units can exert a strong desynchronizing effect in coupled systems, even when its intensity is relatively limited compared to that of the intrinsic activity\cite{teramae2001,boccaletti2002}. Upon injecting noise at a certain location on the ring, its amplitude will thus gradually decrease in the direction of the structural coupling, and will be minimal at the node preceding the injection site. It is therefore possible to use externally-generated noise as a means of inducing localized desynchronization.\\
Experiments were performed setting $G_6=0.12$, $G_{\textrm{i}}=-1.56$, $K_1=0.121\ \mu\textrm{s}^{-1}$ and $G_{\textrm{e}}=10$, introducing Gaussian white noise from an external arbitrary waveform generator (HP33120A; Agilent Inc., Santa Clara CA, USA), with $\sigma_\textrm{noise}\approx300\textrm{ mV}$ and $\sigma_\textrm{signal}/\sigma_\textrm{noise}\approx1.5$. For each noise injection site, synchronization was averaged over 4 runs and, as in Subsection \ref{basics2}, the recorded time-series have length $l=\textrm{65,536 points}$ and were sampled at 8-bit and $f_\textrm{s}=100\textrm{ kSa/s}$.\\
As expected, the relative noise intensity rapidly decreased over few steps after the injection site (Fig. \ref{fig:fig4}a). Desynchronization manifested itself primarily through obliteration of remote synchronization between the nodes before and after the injection site, a result which was well-evident both when averaging the effect over all possible injection sites (Fig. \ref{fig:fig4}b), and when considering individual examples (Figs. \ref{fig:fig4}c,d). Since the difference was primarily confined to the nodes in the vicinity of the injection site, this result provides an initial indication that the observed form of remote synchronization is a ``local'' effect. In other words, it does not reflect an emergent property of the collective dynamics of the ring as a whole. Contrariwise, it represents some sort of ``transformation'' which is at least partially reversed after a certain number of steps from a starting towards an end node. As will become clearer in Section \ref{interactionbands}, this has to do with interaction between the frequency sidebands, involving demodulation and interference effects.\\
Notably, in some cases such as the example shown in Fig. \ref{fig:fig4}d, noise injection had the paradoxical effect of enhancing remote synchronization away from the injection site. This provides a compelling demonstration of the known possibility for weak additive noise to induce and enhance synchronization, in systems with both identical and non-identical units. In this system, small parametric mismatches are present and structured synchronization patterns emerge. Effectively obliterating a node which generates activity that has a primarily repulsive (desynchronizing) effect on other nodes can therefore enhance synchronization in some regions of the network, by means of lowering the energy transfer rate necessary to maintain a given of level of synchrony\cite{boccaletti2002,teramae2004}.
\subsection{Sinc pulse propagation}\label{perturbation2}
On another level, it remains unclear whether the present form of remote synchronization can or not act as a channel for ``hidden'' information propagation, as was previously speculated\cite{minati2015}. In this regard, it is necessary to consider that even when information transfer is rigorously inferred in an observational framework, its detection cannot always be given a cause-effect interpretation unless specific conditions are met\cite{chicharro2012}. Contrariwise, intervening on the system through external perturbations elicits genuine causal effects, and allows their evaluation in terms of \textit{information flow}\cite{pearl2003,lizier2010}. This is exemplified by the hypothetical case of two neurons resected together from brain tissue, which do not generate intrinsic activity but are structurally coupled to each other and therefore capable of exchanging externally-applied stimuli: the information transfer is zero but information flow is not.\\
Addressing this issue, experiments were performed injecting low-amplitude sinc pulses, setting $G_{\textrm{e}}=1$ and a peak amplitude of $\approx500\textrm{ mV}$, corresponding to $\approx1/3$ of the range of intrinsic activity. This allowed studying pulse propagation using a time-locked averaging approach, while minimally disrupting the emergence of remote synchronization (Fig. \ref{fig:fig5}a). We considered sinc pulses having a width of the main lobe of 0.5, 1 and 2 ms, corresponding approximately to 1.5, 3 and 6 times the inverse of the carrier frequency; corresponding run numbers for averaging were 65, 130 and 260. The sampling parameters were the same as indicated above. The peak value was extracted for each window, then compared to 100 random windows by means of a rank-order test, setting significance threshold to $\alpha=0.05$. We observed that, in presence of apparently remote synchronization of intrinsic activity, there was no remote propagation of these pulses: their peak amplitude decayed rapidly and monotonically, disappearing into the baseline after $d=4$ steps (Fig. \ref{fig:fig5}b). Accordingly, even though the sinc pulses were clearly visible in the time-locked average at the injection site (Fig. \ref{fig:fig5}c), they were completely undetectable at the distance corresponding to maximum remote synchronization (Fig. \ref{fig:fig5}d).\\
Recalling that linear Granger causality clearly showed the remote effect (Fig. \ref{fig:fig3}b), these results provide, from the perspective of linear interdependence, a compelling demonstration of decoupling between information flow (measured through intervention, not showing remoteness) and information transfer (measured by linear Granger causality, showing remoteness). The mismatch is clearly related to the fact that a linear model is inadequate for representing the dynamics of this system. Accordingly, the information flow resembled more closely information transfer as given by the non-linear and non-parametric analyses. However, the propagation range for perturbations was seemingly shorter compared to information transfer of intrinsic activity (i.e. information transfer was still visible for $d>4$ steps in Figs. \ref{fig:fig3}c,d).

\begin{figure}
\centering
\includegraphics[width=1\textwidth]{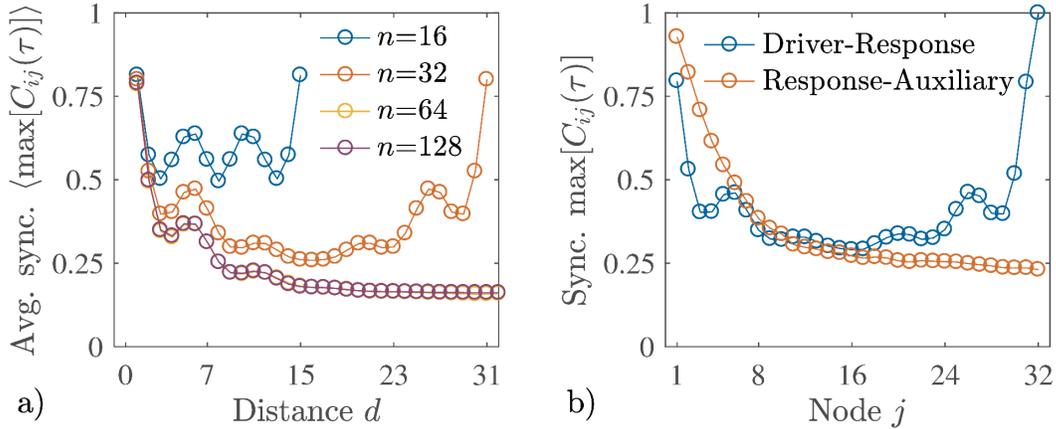}
\cprotect\caption{Effect of ring size, and auxiliary system analysis. a) Average amplitude synchronization as a function of distance along the ring for simulated rings of different size. b) Comparison of driver-response system and response-auxiliary system amplitude synchronization, for driver node $i=32$. See Section \ref{ringsaux} for detailed description.\label{fig:fig6}}
\end{figure}

\section{DIFFERENT-SIZE RINGS AND AUXILIARY SYSTEM}\label{ringsaux}
To garner further confirmation of the local nature of the interactions supporting the emergence of the observed form of remote synchronization, rings of different size $n=16,32,64,128$ (powers of 2 only for convenience, exact choice not relevant) were simulated as in Ref. \citen{minati2015}, each one over 12 runs having random parametric mismatches ($\xi=0.005$, see Subsection \ref{basics1} and Supplementary Section I for detailed description). The average amplitude synchronization was charted as a function of distance and exhibited a peak at $d\approx6$, whose location was remarkably stable across runs and insensitive to ring size (Fig. \ref{fig:fig6}a). Should the remote synchronization effect have been related to a global network phenomenon such as a collective oscillation mode, the location of the peak would plausibly have been sensitive to network size.\\
According to linear cross-correlation and mutual information, synchronization decayed non-monotonically, and a similar pattern was observed for linear Granger causality (see Subsections \ref{basics2} and \ref{causality2}). By contrast, monotonic decay was revealed by non-linear Granger causality and model-free transfer entropy, in agreement with a measure of generalized synchronization used in the initial study  (see Subsection \ref{causality3})\cite{minati2015}. These results indicate that intermediate nodes may be entrained according to a complicated, generalized relationship between trajectories $\mathbf{y}(t)=\mathbf{\Phi}(\mathbf{x}(t))$, where $\mathbf{x}(t)$ and $\mathbf{y}(t)$ are the state vectors of the driver and response systems. On the other hand, more distant nodes revert towards following a simple relationship, detectable also on scalar time-series of the individual state variables, i.e.,  $y(t)\propto x(t)$\cite{boccaletti2002}.\\
This hypothesis was explicitly tested by means of the auxiliary system approach, which involves simultaneously driving a response system and a replica (auxiliary) system, and evaluating the synchronization between the latter two. Effectively, the replica system is used to ``unfold'' a complicated entrainment relationship, allowing its detection with traditional measures such as phase coherence and correlation\cite{Abarbanel1996}.\\
Given a ring of $n=32$ nodes and assuming node $i=n$ as driver, an auxiliary system was constructed as chain of $n$ nodes, having parameters identical to those in the ring but initial conditions randomized to a factor of $100\xi$ (arbitrarily chosen to implement a large spread, not critical). Linear cross-correlation was evaluated between each node in the ring (response system) and the corresponding node in the chain (auxiliary system), and averaged over 100 runs. The intensity of synchronization between the response and auxiliary systems decayed monotonically: it was substantially larger than the synchronization between the driver and response systems for intermediate distances (nodes $j=2\dots5$), and afterwards decayed to comparable values for longer distances $j>5$) (Fig. \ref{fig:fig6}b).\\
This result provides ultimate confirmation that, rather than truly remote entrainment, with increasing distance there is a partially reversible transition between a simple relationship $y(t)\propto x(t)$ and a more complicated one $\mathbf{y}(t)=\mathbf{\Phi}(\mathbf{x}(t))$. This situation recalls the behavior of Mackey-Glass oscillators introduced in Subsection \ref{basics2}, which can sustain generalized synchronization not visible to linear correlation and mutual information\cite{soriano2012}. 

\begin{figure*}
\centering
\includegraphics[width=0.85\textwidth]{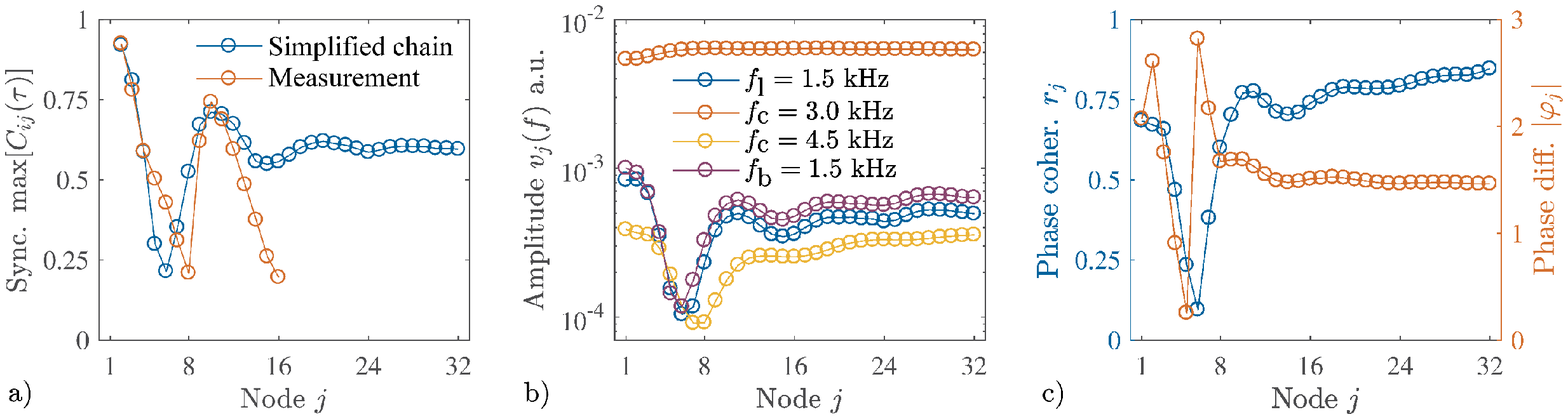}
\cprotect\caption{Simulated chain model. a) Amplitude synchronization measured experimentally along a ring segment (orange, calculated with respect to node $i=1$ representing the input experimental time-series and truncated due to noise injection at node $j=17$, see Fig. \ref{fig:fig4}d), and corresponding simulated results from the simplified chain model (blue). b) Corresponding Fourier amplitudes of the predominant peak ($f_\textrm{c}$, orange, raw signal), lower and higher sidebands ($f_\textrm{l}$ and $f_\textrm{h}$, blue and yellow, raw signal), and demodulated baseband ($f_\textrm{b}$, purple, signal envelope). c) Corresponding phase coherence and absolute phase shift between the lower sideband and demodulated baseband signals within each node $j$. See Subsections \ref{simplmodel1} and \ref{interactionbands2} for detailed description.\label{fig:fig7}}
\end{figure*}

\begin{figure}
\centering
\includegraphics[width=\textwidth]{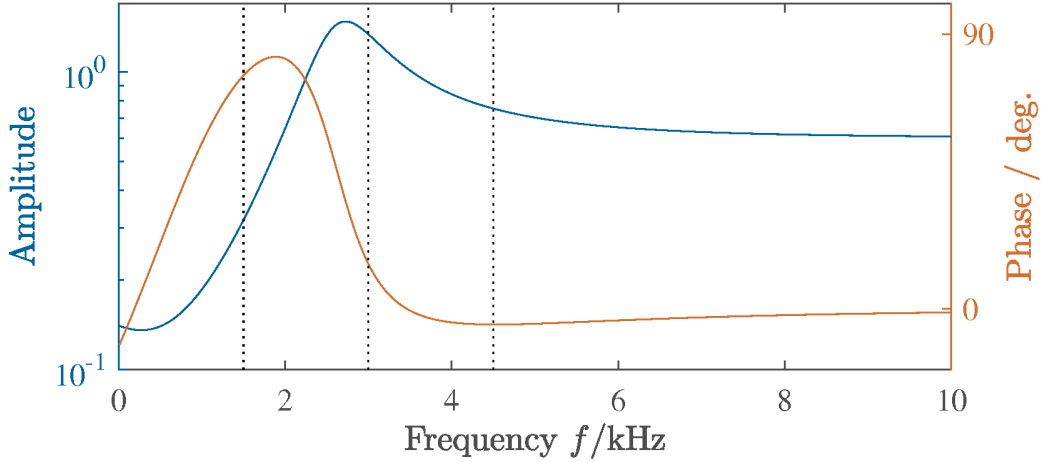}
\cprotect\caption{Amplitude and phase responses of each node according to the simplified model in Eq. (12), rendered linear time-invariant by disregarding the integrator and associated non-linearity, i.e. setting $G_3=0$. Dashed lines denote the locations of $f_{\textrm{l}}=1.5\textrm{ kHz}$, $f_{\textrm{c}}=3.0\textrm{ kHz}$ and $f_{\textrm{h}}=4.5\textrm{ kHz}$. See Subsection \ref{interactionbands2} for detailed description.\label{fig:fig8}}
\end{figure}

\section{SIMPLIFIED CHAIN MODEL}\label{simplmodel}
\subsection{Model specification}\label{simplmodel1}
Further insight into the mechanism underpinning the apparent remote synchronization can be obtained by considering a heavily simplified model, specified as follows.\\ 
First, an open network is considered in the form of a chain, similarly to the auxiliary system. This change removes the possibility of effects related to a cyclic topology. In this case, at the beginning of the chain an experimental signal is supplied as a source of predictive information. The signal from node $j=1$ in Fig. \ref{fig:fig4}d was chosen because this experiment yielded the most pronounced observation of remote synchronization, but the choice is not critical.\\
Second, two dynamical equations are removed: the stage generating $v_6$ is replaced with a simple summer, and the integrator generating $v_5$ is removed altogether. This reduces the order of the system from six to four.\\
Third, the non-linearity $\Gamma(x,y)$ is removed for all voltages except $v_3$. This allows demonstrating that a single instance of the non-linearity is sufficient to reproduce the phenomenon.\\ 
Fourth, the parameters are set identically across all nodes, i.e. $\xi=0$. As discussed below, this enables concluding that generation of the seemingly non-monotonic synchronization decay is not in itself related to the parametric heterogeneities.\\
These simplifications provide insight into which system elements are essential for reproducing the phenomenon. Moreover, as illustrated below, they reveal the possibility of reproducing it without even generating self-sustained oscillations. Having implemented these simplifications, one can rewrite the system equations as 
\begin{empheq}[left=\empheqlbrace]{align}
\frac{\textrm{d}v_{1}}{\textrm{d}t}&=2\pi F(G_4v_4-v_1)\nonumber\\
\frac{\textrm{d}v_{2}}{\textrm{d}t}&=2\pi F(G_1v_{\textrm{o}}-v_2)\nonumber\\
\frac{\textrm{d}v_{3}}{\textrm{d}t}&=\Gamma(Kv_{\textrm{o}},v_3)\\
\frac{\textrm{d}v_{4}}{\textrm{d}t}&=2\pi F(G_2v_2+G_3v_3-v_4)\nonumber
\end{empheq}
where $v_{\textrm{o}}=G_5v_1+G_{\textrm{i}}v_{\textrm{i}}$. To reproduce remote synchronization, the parameters had to be slightly adjusted with a manual procedure, which yielded $G_1=-3.59$, $G_2=-2.82$, $G_3=-0.884$, $G_4=-2.94$, $G_5=0.114$, $G_{\textrm{i}}=0.604$, $F=2\textrm{ kHz}$, $K=0.091\ \mu\textrm{s}^{-1}$ and $V_{\textrm{s}}=4.7\textrm{ V}$. This model closely recovered the non-monotonic variation of amplitude synchronization with distance from the first node observed experimentally, dipping at node $j=6$ and subsequently peaking at node $j=10$ (Fig. \ref{fig:fig7}a).
\subsection{Conditional self-entropy}\label{simplmodel2}
A difference with respect to the experimental system stands out: in simulations of the simplified model, after the initial spatial transient corresponding to the remote synchronization effect, the level of amplitude synchronization with respect to the initial signal stabilized at a relatively high level, i.e. $\max[C(\tau)]\approx0.6$. By contrast, in the experimental system the synchronization rapidly decayed towards $\max[C(\tau)]\approx0.2$ at $d\approx16$.\\
To understand the underlying motivation, we next calculated conditional self-entropy in the simplified chain model and compared it to the full ring. Conditional self-entropy is akin to transfer entropy, but it is obtained by swapping the roles of the past states of the target and driver. It therefore quantifies the amount of information which is ``transferred internally'' within the target node, computing the mutual information between the present and past target dynamics conditioned to the past dynamics of the driver. Conditioning on the source past removes the redundancy between the source and target past, and adds the synergies between them.\\
The conditional self-entropy differs from the self-entropy in the fact that conditioning on the driver activity has the effect of removing the ``input-driven storage''\cite{obst2013}, and this precise effect allows us to highlight the presence of internal dynamics within a node. The conditional self-entropy can therefore be viewed as a measure of the extent to which a node is an ``active'' signal generator as opposed to a ``passive'' entity simply relaying the input signal, albeit with a potentially non-linear transformation\cite{faes2015}.\\
We observed that the conditional self-entropy in the simplified chain model peaked at node $j=4$ with $S_{4|3}=0.09$ and thereafter gradually decreased with small fluctuations along the chain until $S_{32|31}=0.04$, being overall substantially lower compared to the full ring in the chaotic state, i.e. $0.04\pm0.02$ vs. $0.38\pm0.06$, $p<0.001$.
\subsection{Interpretation}\label{simplmodel3}
The above result reveals that although the simplified model can closely reproduce the remote synchronization effect when fed an experimental time-series as a source of predictive information at the beginning of the chain, crucially its nodes are, at least to a large extent, not ``active'' signal generators. In other words, they do not generate predictive information independently of that received from the other connected nodes.\\
It is for this reason that, after the initial transient, the level of synchronization with respect to the initial time-series remains high. In the full system, each node acts as a source of predictive information and thus generates an own signal; as this is overlapped to the input signal, after the initial spatial transient a synchronization decay is observed owing to diffusion. In turn, this finding indicates that in the full system the non-linearity plays two distinct roles. One is generative, namely supporting chaotic dynamics. The other, as will be clarified below, consists of enabling energy transfer between the frequency bands.\\
In this simplified model, the non-linearity largely or solely plays the latter role, and it is for this reason that parametric mismatches are not necessary. On the contrary, as demonstrated in Supplementary Section I, in the full model it is necessary to have $\xi>0$. This is because when each node generates an own signal, parametric heterogeneities effectively increase the energy transfer rate necessary to maintain synchronization. This is a form of ``repulsion'' which, as also observed in other systems, prevents entering the globally-synchronized state\cite{boccaletti2002,ottbook}, and in this case allows the observed form of remote synchronization to emerge.

\begin{figure*}
\floatbox[{\capbeside\thisfloatsetup{capbesideposition={right,center},capbesidewidth=5cm}}]{figure}[\FBwidth]
{\caption{Representative time-series and spectra from the simplified model. a) and b) Instantaneous amplitudes, comparing the activity in node $j=1$ to that in nodes $j=6,\ 10$, representing the synchronization dip and the subsequent recovery. Mean voltage has been subtracted. c) and d) Corresponding frequency spectra for the raw signals and envelopes. See Subsections \ref{simplmodel1}, \ref{interactionbands1} and \ref{interactionbands2} for detailed descriptions.\label{fig:fig9}}}
{\includegraphics[width=0.55\textwidth]{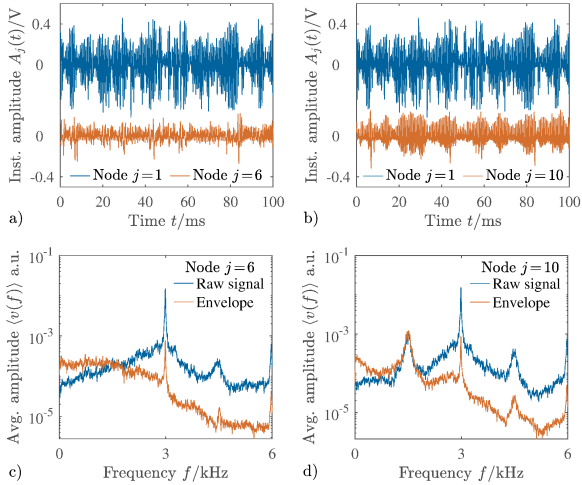}}
\end{figure*}

\section{DEMODULATION AND INTERFERENCE}\label{interactionbands}
\subsection{Demodulation}\label{interactionbands1}
By means of noise perturbation experiments, simulations on rings of different size and the simplified chain model, it has been demonstrated that the apparent remote synchronization is underpinned by local interactions. It shall now be revealed that the underlying phenomenon consists of a form of cross-frequency exchange.\\
In simulations of the simplified model, the central, lower and higher peaks introduced in Subsection \ref{basics2} are found at $f_{\textrm{c}}=3.0\textrm{ kHz}$, $f_{\textrm{l}}=1.5\textrm{ kHz}$ and $f_{\textrm{h}}=4.5\textrm{ kHz}$ respectively. The central peak is predominant and reflects oscillation periodicity. As detailed in Supplementary Section I, in this system the lower peak at frequency $f_{\textrm{l}}=f_{\textrm{c}}/2$ hallmarks anti-persistent dynamics, stemming from an underlying period-doubling bifurcation. On the other hand, the higher peak corresponds to its image frequency with respect to the central peak, $f_{\textrm{h}}=f_{\textrm{l}}+f_{\textrm{c}}$.\\
Since the central peak is predominant and the weaker sidebands are symmetric with respect to it, the spectrum recalls that of standard amplitude modulation (AM) with carrier frequency $f_{\textrm{c}}$ and baseband signal at frequency $f_{\textrm{b}}=f_{\textrm{c}}-f_{\textrm{l}}=f_{\textrm{h}}-f_{\textrm{c}}$. However, unlike typical applications of AM techniques, the relationship $f_{\textrm{h}}=f_{\textrm{l}}+f_{\textrm{c}}$ is simultaneously verified. This requires that the lower sideband and the baseband are spectrally coincident, which is only possible if $f_{\textrm{l}}=f_{\textrm{b}}=f_{\textrm{c}}/2$. Such situation arises quite commonly in chaotic systems that have anti-persistent dynamics. However it is not found, for example, whenever chaoticity follows non-linear interaction between oscillations at unrelated frequencies. Examples of both scenarios can be found among transistor-based chaotic oscillator circuits\cite{ottbook,minati2017}.\\
This specific spectral configuration is however plausibly not a necessary condition for the emergence of forms of remote synchronization such as the present one, because interference between preexisting and demodulated baseband signal as described below could in principle also arise with $f_{\textrm{l}}\neq f_{\textrm{b}}$. It is also not a sufficient condition, because in the present system it is observed over wide ranges of parameter settings, including regions without remote synchronization (data not shown). Notably, amplitude modulation has previously been considered as a possible means of information transmission in chaotic networks, however in a different framework\cite{cessac2006}.\\
In the simplified chain model described in Section \ref{simplmodel}, while the amplitude of the central peak remained relatively constant along the chain, the amplitudes of the sidebands underwent large spatial fluctuations. Namely, one observes that the synchronization dip at node $j=6$ (Fig. \ref{fig:fig7}a) is accompanied with partial cancellation particularly of the lower sideband, followed by recovery showing damped spatial oscillation, and resembling a diffraction pattern (Fig. \ref{fig:fig7}b). It shall now be clarified that this association reflects an interference effect.
\subsection{Interference}\label{interactionbands2}
To elucidate the mechanism underlying interference, it is necessary to examine more closely the properties of each node (Eq. (1) and Fig. \ref{fig:fig1}a for the full system; Eq. (12) for the simplified model).\\
In particular, it is useful to consider the response of the simplified model, further reduced and rendered linear time-invariant by neglecting the integrator and associated saturation non-linearity, i.e. setting $G_3=0$. Due to the ring structure, formed of the three composite stages implementing gain and low-pass filtering and realizing a resonator, and summation of their output to the input from the previous node, each node has frequency and phase responses which resemble a peaking filter at low frequencies and a simple pass-through at high frequencies. For the chosen parameter settings, this arrangement emphasizes the higher sideband relative to the lower one, which it significantly attenuates and dephases (Fig. \ref{fig:fig8}).\\
Due to presence of the integrator(s) and associated saturation non-linearity, another signal component in the frequency range of the lower sideband is concomitantly generated. In particular, because saturation occurs in a highly asymmetric manner, the non-linearity effectively realizes an envelope detector, resembling the demodulation mechanism found in early AM radio receivers. In fact, further simplifying the non-linearity so that it includes only negative saturation
\begin{eqnarray}
\Gamma\left(x,y\right)=\textrm{R}\left(x\right)-\textrm{R}\left(-x\right)\textrm{H}(V_{\textrm{s}}+y)
\end{eqnarray}
the close overlap between experimental data and the chain model can still be perfectly reproduced (Fig. \ref{fig:fig7}a). Depending on the phase relationship between the low-frequency component preexistent in the input signal and such demodulated signal, constructive or destructive interference can arise. In turn, this also reflects in the higher sideband, but does so to a lesser extent, since the non-linearity has a primarily demodulating action.\\
The fact that in the simplified chain model each node acts as a relay without overlapping an own chaotic signal facilitates demonstrating how the ``hidden'' baseband signal (amplitude fluctuations) can easily be recovered. For example, let us consider a second-order peaking filter having manually-chosen $f_0=4.425\textrm{ kHz}$ and $Q=4.5$, and calculate amplitude synchronization based on the envelope of the signal thus filtered. The dip in synchronization with respect to the initial signal is drastically attenuated, with corresponding dip $\max[C(\tau)]=0.55$ at node $j=8$ as opposed to $\max[C(\tau)]=0.20$ at node $j=6$, rendering the synchronization of more distant nodes no longer remote. To recover the ``hidden'' signal, it is therefore sufficient to perform a linear operation, namely attenuate the lower sideband and emphasize the higher sideband, and then obtain the baseband signal by demodulation, that is, envelope detection. However, the non-linearity is necessary to obtain the non-monotonic spatial relationship underlying remote synchronization.\\
To further clarify the manner in which the synchronization dip appears at a certain distance, it is helpful to consider the spatial variation along the chain of the phase relationship between the lower sideband and the demodulated baseband signals within each node. For explanatory purposes, the latter can simply be recovered via the same peaking filter. Such spatial variation stems from the master-slave coupling architecture, which cascades the responses and dynamics of the nodes.\\
The phase coherence between the lower sideband and the demodulated baseband signals $r_j$ dipped at node $j=6$, corresponding to the point of maximally-destructive interference, then raised again, peaking at node $j=10$. Consideration of the absolute phase shift $|\varphi_j|$ further indicated that the relationship between the lower sideband and demodulated baseband signals rapidly drifted at the beginning of the chain, until a stable phase shift was attained after node $j=8$ (Fig. \ref{fig:fig7}c). The non-monotonic changes of synchronization observed in the simplified chain model, as well as in the full ring, are therefore underpinned by the phase shift between the lower sideband signal and the baseband signal. Such phase shift is initially spatially unstable, and at a certain distance leads to almost-complete cancellation. This distance depends in a non-trivial manner on the system parameters, particularly filter frequency $F$ (data not shown). The carrier and higher sideband are sufficiently preserved, and the low frequency signal is eventually recovered from them (Fig. \ref{fig:fig9}a-d).
\subsection{Effect of instancing additional filters}\label{interactionbands3}
\begin{figure*}
\floatbox[{\capbeside\thisfloatsetup{capbesideposition={right,center},capbesidewidth=5cm}}]{figure}[\FBwidth]
{\caption{Intervention consisting of attenuating either sideband in the simplified model. a) Average frequency spectra of the initial, low- and high-pass filtered signals (blue, yellow, orange). b), c) and d) Effect of inserting the corresponding low- and high-pass filters along the chain model after nodes $j=3,6,8$. See Subsection \ref{interactionbands3} for detailed description.\label{fig:fig10}}}
{\includegraphics[width=0.55\textwidth]{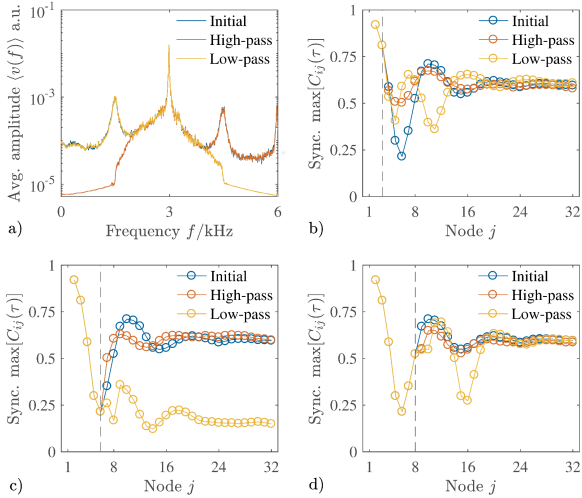}}
\end{figure*}

This account of the remote synchronization mechanism can be corroborated by directly intervening in the simplified chain model, i.e. instancing filters which selectively attenuate either sideband. Additional simulations were therefore performed adding at specified sites either a high-pass filter having $f_{\textrm{stop}}=1.5\textrm{ kHz}$ and $f_{\textrm{pass}}=2.0\textrm{ kHz}$ (only removing the lower sideband), or a low-pass filter having $f_{\textrm{pass}}=4.0\textrm{ kHz}$ and $f_{\textrm{stop}}=4.5\textrm{ kHz}$ (only removing the higher sideband). In both cases, finite impulse response (FIR) equiripple filters with passband ripple 1 dB, stop-band attenuation 40 dB were used (Fig. \ref{fig:fig10}a).\\
First, we instanced the filters before the synchronization dip, after node $j=3$. As expected for an interference effect, removing either sideband, particularly the lower one, attenuated the  dip; furthermore, removing the higher sideband introduced a shallower desynchronization peak further down the chain around nodes 10-12 (Fig. \ref{fig:fig10}b).\\
Second, we instanced the filters at the site corresponding to the synchronization dip, namely after node $j=6$. Here, a striking difference was observed between the effect of removing the lower and the higher sideband: while removing the lower sideband had negligible effect on the recovery of synchronization further down the chain ($0.60\pm0.02$ vs. $0.60\pm0.02$, $p=0.6$), removing the higher sideband effectively obliterated it ($0.20\pm0.04$ vs. $0.60\pm0.02$, $p<0.001$; Fig. \ref{fig:fig10}c).\\
Third, we instanced the filters just two steps further down the chain, after node $j=8$, observing a totally different effect, whereby removing the higher sideband did not impair the eventual recovery of synchronization, but only engendered a new interference dip around node $j=16$ (Fig. \ref{fig:fig10}d).\\
Although partially dependent on the filter choices (data not shown), these interventions confirm the occurrence of significant energy exchange between the frequency sidebands, and subsequent interference. More specifically, they reveal that at the point of synchronization dip, the ``hidden'' synchronization information is conveyed predominantly through the higher sideband and carrier signal.\\
Accordingly, as shown in Supplementary Section II, a linear causality model is actually capable of resolving the underlying information transfer, insofar as the activity of both sidebands is represented in a manner appropriate for the system, in this case through filtering and demodulating the higher sideband. It is in fact the relationship between the frequency sidebands which in Subsection \ref{causality3} could be approximated by the quadratic Granger model, and agnostically captured by the model-free transfer entropy analysis.

\section{DISCUSSION}\label{conclusions}
\subsection{Phenomenon}
To the authors' knowledge, consensus has not yet been reached regarding what phenomena should properly be termed remote synchronization.\\
In the case of systems with parametric heterogeneities, and even more so in presence of large mismatches, the possibility of identical synchronization may be largely precluded\cite{perlikowski2010routes,yanchuk2011variability,d2011role}. The reported instances of remote synchronization in such systems involve a more complicated form of synchronization for the relay nodes, such as a generalized relationship between trajectories. Synchronization appears to be remote due to the usage of synchronization (or causality) measures which are not sufficiently general to represent the entrainment of the relay or intermediate nodes\cite{bergner2012remote,fischer2006zero,gutierrez2013generalized}. Recourse to suitable measures\cite{Chicharro2009} or application of the auxiliary system approach\cite{boccaletti2002,Abarbanel1996} reveals non-remote entrainment, which in the present case is evident as a monotonic decay of synchronization strength along the structural connections. It can therefore be argued that such instances only reflect an apparent form of remote synchronization. The present observations fall into that category.\\
On the other hand, in networks of identical elements remote entrainment cannot simply be resolved by making recourse to generalized synchronization measures: in other words, remoteness is a product of topological features such as network symmetries \cite{nicosia2013remote,zhang2017incoherence}. Such instances represent a qualitatively-different scenario, compared to the observations in systems with parametric heterogeneities such as the present one.\\
There are three aspects of the present system which seem novel and noteworthy.\\
First, the apparently remote synchronization of amplitude fluctuations is supported by generalized synchronization of intermediate nodes, which are only subject to small parametric mismatches. It appeared in simulations for mismatches as small as $\xi\approx10^{-17}$. This represents an important difference compared to previous observations in star networks\cite{bergner2012remote} and complex networks\cite{gambuzza2013analysis}. Whereas in Refs. \citen{bergner2012remote} and \citen{gambuzza2013analysis} synchronization of the relay node(s) is purposefully hampered by the mismatched oscillation frequency, here the synchronization dip emerges due to a demodulation and interference phenomenon that is generated by the system itself. Parametric mismatches in this present system are thus less closely related to the remoteness effect, and primarily needed to avoid entering a globally-synchronized state. It could therefore be argued that in this system the observed form of remote synchronization more closely reflects an emergent property, instead of a designed feature.\\
Second, as more extensively discussed in Ref. \citen{minati2015}, in this system the amplitude fluctuations are not a means of indirectly maintaining phase synchronization, but become themselves synchronized in an apparently remote (i.e., generalized) form.\\
Third, for short distances along the ring, the synchronization relationship has the form $y(t)\propto x(t)$. For intermediate distances, it takes a more complicated form $\mathbf{y}(t)=\mathbf{\Phi}(\mathbf{x}(t))$, and for longer distances it reverts back towards the form $y(t)\propto x(t)$, before eventually vanishing. It therefore represents a partially-reversible ``transformation''. This pattern is produced in a staggered arrangement, clearly visible in the synchronization and causality matrices as ``diagonal lines''. This peculiar result reflects that fact that, being the parametric mismatches small, the nodes are effectively equivalent. Each node therefore acts both as a generator of predictive information and as a relay station, wherein the sideband and baseband signals of multiple preceding nodes at different distances overlap.
\subsection{Generality}
Finally, the generality of the phenomenon needs to be discussed in relation to both the network topology and the unit dynamics, which represent two separate and independent aspects of a system.\\
First, as far as topology is concerned, we have showed that the observed form of remote synchronization does not require a cyclic topology and may therefore arise in any network containing a chain motif. Since this effectively corresponds to the totality of known complex networks, future work should extend the present findings to complex topologies, thus addressing the potential relevance of this phenomenon to collective behaviors in more naturalistic scenarios, as previously done for remote synchronization in Stuart-Landau systems\cite{gambuzza2016inhomogeneity}. However, the spatial phase gradient leading to interference at a specific location can only be maintained in the context of unidirectional coupling, and would be dissipated under diffusive coupling, which renders the observed phenomenon specifically pertinent to directed networks.\\
Second, with regards to the dynamics of the units, as observed in other physical systems such as plasmas\cite{sodha1968}, any suitable non-linearity can have a demodulating effect. There is thus no special relevance of the saturation non-linearity present at the integrator output(s). However, the spectral features are crucial in determining whether or not an interference effect can arise, and this limits the situations under which the described mechanism may apply.\\
The most restrictive requirement is that the spectrum of chaotic activity should not be broadband, but contain distinct peaks, namely be close to quasiperiodicity. For broadband spectra, cancellation due to interference is not plausible. It is known that power spectral features are more closely related to the mixing properties of strange attractors than to the system characteristics. Depending on the region, chaotic systems can generate either broadband signals or power spectra with distinct peaks, as exemplified by the band and funnel attractors in the R{\"o}ssler system \cite{farmer1980}. The requirement for distinct peaks therefore does not appear, in itself, to exclude a-priori observing the demodulation and interference effect in any chaotic system.\\
There are, however, obvious additional requirements on the locations of the peaks. Namely, for demodulation and subsequent interference to occur there is a need for amplitude fluctuations to be driven by a baseband signal, which can be also demodulated from other frequencies, possibly with a phase difference. One suitable scenario is a spectrum containing distinct peaks at harmonic frequencies $\Omega,\ 2\Omega\ldots$ etc. These can be generated, for example, by period-doubling bifurcations, but this is certainly not the only mechanism\cite{ottbook}. The effect can also arise whenever there is a baseband signal at frequency any $\Omega$, together with at least two higher frequency components at $\Omega^\prime$ and $\Omega^\prime+\Omega$, where $\Omega^\prime=(1+k)\Omega$ and any $k>0$. Such situation can be implemented by design, or may reflect some emergent property. Evidently, the effect cannot arise if the spectrum only features peaks located at unrelated frequencies, as may be found in some cases when chaos arises via the quasiperiodicity route\cite{minati2017,ottbook}.\\
More generally, even though putatively self-organized critical systems such as the brain tend to generate broadband signals, this does not exclude the possibility of observing narrower spectra, and consequently interference, at the microscopic scale \cite{he2014}. The observed interference effect recalls two-tone interference in hearing, in particular its manifestation as two-tone suppression \cite{julicher2001}. Furthermore, amplitude modulation and interference effects have previously surfaced in models of asynchronous neuronal populations closely fitting extracellular recordings \cite{diaz2007}. The importance of cross-frequency coupling effects in neural systems is only recently being considered, and it is plausible that complex pattern formation mechanisms such as the one observed in the present study may also underpin the mismatch observed between entrainment of spontaneous activity and responses to external perturbations\cite{jirsa2013,bortoletto2015}.
\section*{SUPPLEMENTARY MATERIAL}
See supplementary material for detailed results on the chaotic transitions and effect of parametric mismatches, and on the revised Granger causality model.
\section*{ACKNOWLEDGMENTS}
All experimental activities were self-funded by L.M. personally and conducted on own premises.
\bibliographystyle{aip}

\end{document}